\documentclass[12pt]{article} 
\usepackage[T1]{fontenc}
\usepackage{amsmath,amssymb,amsfonts,amsthm}
\usepackage{graphicx}
\usepackage{enumitem}
\usepackage{bm}
\usepackage{rotating}
\usepackage{hyperref}
\hypersetup{
    plainpages=false,       % needed if Roman numbers in frontpages
    unicode=true,          % non-Latin characters in Acrobat’s bookmarks
    pdftoolbar=true,        % show Acrobat’s toolbar?
    pdfmenubar=true,        % show Acrobat’s menu?
    pdffitwindow=false,     % window fit to page when opened
    pdfstartview={FitH},    % fits the width of the page to the window
    pdfnewwindow=true,      % links in new window
    colorlinks=true,        % false: boxed links; true: colored links
    linkcolor=RoyalPurple,         % color of internal links
    citecolor=BrickRed,        % color of links to bibliography
    filecolor=magenta,      % color of file links
    urlcolor=RoyalBlue           % color of external links
}
\usepackage{pbox}
\usepackage[dvipsnames]{xcolor}
\usepackage{array}
\usepackage{physics}
\usepackage{dsfont}
\usepackage{mathtools}
\usepackage{leftidx}
\usepackage[normalem]{ulem}
\usepackage{braket}
\usepackage{tensor}
\usepackage{mathrsfs}
\usepackage{float}
\usepackage{dsfont}
\usepackage[margin=1in]{geometry}
\usepackage[title]{appendix}
\usepackage{tcolorbox}
\usepackage{mathpazo} % closest to Palatino
\usepackage{setspace}
\usepackage{tikz}
\usetikzlibrary{arrows.meta,decorations.pathmorphing,math}

\renewcommand{\bar}{\overline}

\newcommand{\skri}{\mathscr{I}}

\newcommand{\sx}{\mathsf{x}}

\newcommand{\rr}[1]{\left(#1\right)}
\newcommand{\bx}{{\bm{x}}}

\newcommand{\openone}{\mathds{1}}

\newcommand{\M}{\mathcal{M}}

\newcommand{\CS}{C^\infty_c(\M)}

\newcommand{\rao}{\hat{\rho}_{\textsc{a},0}}
\newcommand{\rbo}{\hat{\rho}_{\textsc{b},0}}

\newcommand{\rof}{\hat{\rho}_{\phi,0}}

\begin{document}

\title{Fuzzy spacetime: fundamental limits of quantum-optical holographic bulk reconstruction}

\author{Erickson Tjoa$^{1,2}$\footnote{Corresponding author: \href{mailto:erickson.tjoa@uwaterloo.ca}{erickson.tjoa@uwaterloo.ca}}\\
$^1$Department of Physics and Astronomy, University of Waterloo\\
$^2$Institute for Quantum Computing, University of Waterloo}
%\affiliation{Department of Physics and Astronomy, University of Waterloo, Waterloo, Ontario, N2L 3G1, Canada}
%\affiliation{Institute for Quantum Computing, University of Waterloo, Waterloo, Ontario, N2L 3G1, Canada}

\date{April 19, 2023}
%\email{e2tjoa@uwaterloo.ca}
%\affiliation{Department of Physics and Astronomy, University of Waterloo, Waterloo, Ontario, N2L 3G1, Canada}
%\affiliation{Institute for Quantum Computing, University of Waterloo, Waterloo, Ontario, N2L 3G1, Canada}

%\author{Finnian Gray \footnote{\href{mailto:fgray@perimeterinstitute.ca}{fgray@perimeterinstitute.ca}}}%
%\affiliation{Department of Physics and Astronomy, University of Waterloo, Waterloo, Ontario, N2L 3G1, Canada}
%\affiliation{Perimeter Institute for Theoretical Physics, Waterloo, Ontario N2L 2Y5, Canada}

%\date{March 30, 2023}
\date{\today}

\maketitle
\flushbottom

\begin{abstract}

\noindent {In this Essay we construct a concrete, \textit{non-perturbative} realization of metric reconstruction using quantum-optical model of particle detectors in relativistic quantum information.} The non-perturbative approach allows us to realize a version of ``short-distance physics corresponds to poor statistics'' idea by Kempf which occurs way above the Planck scale before one reaches the quantum-gravitational regime. {In particular, the ``fuzziness'' of spacetime that arise from operational measurement protocols} can be given a holographic dual interpretation using bulk-to-boundary correspondence between scalar correlators in asymptotically flat spacetimes. {The holographic interpretation necessitates imperfect metric reconstruction even in principle due to the universality of future null infinity.}

%we argue, using the ideas in   that there are fundamental limits imposed by quantum theory and thermodynamics on  spacetime metric reconstruction using localized quantum mechanical probes: the ``fuzziness'' of spacetime that arise from operational measurement protocols is already present before one reaches the quantum-gravitational regime. We do this 
%We provide some discussion on what this implies for approaches of quantum gravity. 

\begin{center}
    \emph{“Essay written for the Gravity Research Foundation 2023 Awards for Essays on Gravitation.”}
\end{center}
\end{abstract}

\clearpage 
%\section{Introduction}

Attempts to unify quantum theory and general relativity have lasted for several decades and multiple approaches have been pursued. Approaches range from considering some form of discreteness of spacetime such as causal sets and loop quantum gravity, to more field-theoretic approaches such as string theory and holographic duality. One thing is clear: from each approach one learns various insights as to why gravity is so special among all the fundamental forces of nature. 

One particularly recent proposal involves abandoning the notion that spacetime is merely a stage (though itself dynamical) on which actors (namely, matter content) move \cite{Kempf2021replace}. The basic idea is inspired from quantum field theory (QFT) in curved spacetimes. Since quantum fields have vacuum fluctuations at short distances and correlations decay at large distances, it is possible in principle that measurements of $n$-point correlation functions of the quantum field furnish us with a way to reconstruct the metric. Indeed, it was shown that the metric tensor on a (pseudo-)Riemannian manifolds can be reconstructed from the Feynman propagator $G_F(\sx,\sx')=\braket{\mathcal{T}\hat{\phi}(\sx)\hat{\phi}(\sx')}$ of a scalar field $\hat{\phi}$, where $\mathcal{T}$ is the time-ordering symbol \cite{Kempf2016curvature}. Soon after, two clever refinements of this proposal appeared.

First, it was shown in \cite{Pipo2018direct} that one can in principle reconstruct the vacuum Wightman two-point functions $\mathsf{W}(\sx,\sx') = \braket{0|\hat{\phi}(\sx)\hat{\phi}(\sx')|0}$ from the density matrix of a quantum-mechanical ``detector'' that couples to the field via dipole-type interaction in quantum optics. This was quickly generalized to curved spacetimes in \cite{perche2021geometry}, which also explains why and how the construction works at all. The key has to do with the fact that all physically reasonable quantum field states are required to be \textit{Hadamard states} \cite{Radzikowski1996microlocal,KayWald1991theorems,Khavkhine2015AQFT,wald1994quantum,fewster2019algebraic}. The Hadamard property is the leading short-distance (ultraviolet) behaviour of the Wightman two-point functions and reflects in some sense the local flatness property of general relativity and the finiteness of fluctuations of observables in QFT.  

In this Essay we will combine the insights in \cite{Kempf2021replace,Pipo2018direct,perche2021geometry} with \textit{non-perturbative} calculation of relativistic quantum channels in \cite{Landulfo2016magnus1,Landulfo2021cost,tjoa2022channel} to perform the metric reconstruction ``quantum-optically''. %\sout{Our calculation puts greater emphasis on the (quantum) information-theoretic flavor and} 
The resulting calculation has three benefits: (1) we will fully realize concretely the notion that ``the Planck scale might appear as a regime of poor statistics'' \cite{Kempf2021replace}; (2) we make explicit how the spatial extent and duration of the interaction, as well as thermodynamic considerations, provide obstructions to perfect metric reconstruction; (3) it is compatible with \textit{modest holography} proposal \cite{tjoa2022modest}, in that we can reconstruct the metric from the null boundary of asymptotically flat spacetime. {In particular, the modest holography  \textit{necessitates} imperfect holographic bulk metric reconstruction even in principle.} %\tcb{The quantum information theory comes in when we assess the measurement statistics.}

%\section{Setup}

Consider two observers Alice and Bob, each carrying an Unruh-DeWitt (UDW) detector \cite{Unruh1979evaporation,DeWitt1979} in a globally hyperbolic spacetime $(\M,g_{ab})$. Each UDW detector is a two-level system (``a qubit'') with free Hamiltonian $\hat{\mathfrak{h}}_j = \frac{\Omega}{2}\hat{\sigma}^z_j$, where $j=A,B$ and $\hat{\sigma}^z_j$ is the Pauli-$Z$ operator. Each qubit has an energy gap $\Omega$. The center of mass of each detector moves along the worldline $\sx_j(\tau_j)$ parametrized by proper time $\tau_j$. The qubit-field interaction Hamiltonian is prescribed as an operator-valued four-form (in interaction picture) \cite{Tales2020GRQO}
\begin{align}
    \hat{h}_{I,j} &= \dd V\,f_j(\sx)\hat\sigma^x_j(\tau_j(\sx))\otimes\hat\phi(\sx)\,,
    \label{eq: UDW-covariant}
\end{align}
where $\dd V = \dd^4\sx \sqrt{-g}$ is the invariant volume element in $\M$;  $f_j(\sx)\in\CS$ prescribes the interaction region between detector $j$ and the field. Eq.~\eqref{eq: UDW-covariant} is nothing but the covariantly reformulated, scalar version of dipole interaction $\hat{\mathbf{d}}\hat{\cdot\mathbf{E}}$ in quantum optics \cite{Lopp2021deloc}. The total unitary time-evolution for the detector-field system is given by the time-ordered exponential (in interaction picture) \cite{Tales2020GRQO,Bruno2021broken}
\begin{align}
    \hat{U} = \mathcal{T}_t\exp\left[-i  \int_\M \dd V \,\hat{h}_{I,A}+\hat{h}_{I,B} \right]\,.
    \label{eq: total-unitary}
\end{align}
Since the spacetime is globally hyperbolic, we can take the time ordering to be with respect to the global time function $t$. Without loss of generality we can set $t(\tau_{j,0}) = \tau_{j,0}=0$. 

Suppose that the detectors are turned on very rapidly, effectively at a single instant in time in their own rest frames: in terms of Fermi normal coordinates $\bar{\sx}\coloneqq (\tau,\bar\bx)$ we have
\begin{align}
    f_j(\tau,\bar{\bx}) = \lambda_j\delta(\tau-\tau_{j,0})F(\bar{\bx})\,,
    \label{eq: delta-interaction}
\end{align}
where $\lambda_j$ is the coupling strength and $F$ defines the spatial profile (``atomic orbitals'') of the detector in its own rest frame. Under this assumption, the unitary \eqref{eq: total-unitary} can be shown to factorize into $\hat{U} = \hat{U}_\textsc{b}\hat{U}_\textsc{a}$ \cite{Simidzija2018nogo,tjoa2022channel}, where
\begin{align}
    \hat{U}_j &= \openone_j\otimes \hat{C}_j - i \hat{X}_j\otimes \hat{S}_j\,,\qquad \hat{X}_j\coloneqq \hat{\sigma}^x_j(\tau_{j,0})\,,\,\hat{C}_j \coloneqq \cos\hat{\phi}(f_j)\,,\,\hat{S}_j\coloneqq \sin\hat{\phi}(f_j)\,.
\end{align}
For compactly supported\footnote{Strictly speaking Eq.~\eqref{eq: delta-interaction} does not define an element of $C^\infty_c(\M)$, but it can be approximated arbitrarily well by elements of $C^\infty_c(\M)$. We could also use gapless models \cite{Landulfo2021cost} to avoid this technicality and obtain essentially analogous results.} $f\in C^\infty_c(\M)$,  $\hat{\phi}(f)$ is called the \textit{smeared field operator} \cite{Khavkhine2015AQFT,fewster2019algebraic,wald1994quantum}
\begin{align}
    \hat{\phi}(f) \coloneqq \int_{\M}\dd V\,f(\sx)\hat{\phi}(\sx)\,,
\end{align}
which is the basic object in scalar QFT. These operators obey the canonical commutation relation (CCR) $[\hat{\phi}(f),\hat{\phi}(g)]=i  E(f,g)\openone$, where $E(f,g)$ is the \textit{smeared causal propagator}\footnote{Causal propagator of the Klein-Gordon wave equation $(\nabla_a\nabla^a - m^2-\xi R)\phi=0$ is nothing but the \textit{advanced-minus-retarded propagator} \cite{fewster2019algebraic,Khavkhine2015AQFT,wald1994quantum}.}.

Suppose that the initial state is an uncorrelated state $\hat{\rho}_0 = \hat{\rho}_{\textsc{ab},0}\otimes\rof$. Then the final state of both detectors is obtained by partial trace over the field, giving rise to quantum channel $\mathcal{E}$ acting on density operators of the detector:
\begin{align}
    \hat{\rho}_{\textsc{ab}} &= \mathcal{E}(\hat{\rho}_{\textsc{ab},0}) = \tr_\phi\bigr(\hat{U}(\hat{\rho}_{\textsc{ab},0}\otimes\rof)\hat{U}^\dagger\bigr)\,.
\end{align}
For simplicity, assume that initially Alice and Bob are uncorrelated so that $\hat{\rho}_{\textsc{ab},0}=\rao\otimes\rbo$, and that the field is in a quasifree state \cite{Khavkhine2015AQFT,KayWald1991theorems,wald1994quantum,fewster2019algebraic} (which includes vacuum and thermal states). Then we have

\begin{equation}
\begin{aligned}
    \hat{\rho}_\textsc{ab}&= \rbo\otimes\left(\rao\gamma_{cccc} + \hat{X}_\textsc{a}\rao\hat{X}_\textsc{a}\gamma_{sccs}\right) + \hat{X}_\textsc{b}\rbo \hat{X}_\textsc{b}\otimes\rr{\rao\gamma_{cssc} + \hat{X}_\textsc{a}\rao\hat{X}_\textsc{a}\gamma_{ssss}} \\
    &+ i  \rbo\hat{X}_\textsc{b}\otimes \rr{i \rao\hat{X}_\textsc{a}\gamma_{sscc} - i  \hat{X}_\textsc{a}\rao\gamma_{cscs}} - i  \hat{X}_\textsc{b}\rbo\otimes \rr{i  \rao \hat{X}_\textsc{a}\gamma_{scsc}-i \hat{X}_\textsc{a}\rao\gamma_{ccss}}\,.
\end{aligned}
\end{equation}
The coefficients are computed as follows: $\gamma_{cccc} = \text{tr}(\rof C_\textsc{a}C_{\textsc{b}}C_{\textsc{b}}C_{\textsc{a}})$ in the ordering $\textsc{ABBA}$, which can be done quickly exploiting the structure of the Weyl algebra (see, e.g., \cite{tjoa2022channel,tjoa2022fermi}). 

Suppose that Alice and Bob initially only has access to local thermal states  $\hat{\rho}_{\textsc{ab},0} = \hat{\rho}_{\textsc{a}}(\beta_\textsc{a})
\otimes  \hat{\rho}_{\textsc{a}}(\beta_\textsc{b})$ with inverse local temperatures $\beta_j$. In the eigenbasis of $\hat{\sigma}^z$, we have
\begin{align}
    \hat{\rho}_{\textsc{a}}(\beta_\textsc{a}) = \begin{bmatrix}
    a_+ & 0 \\ 0 & a_-
    \end{bmatrix}\,,\qquad 
    \hat{\rho}_{\textsc{b}}(\beta_\textsc{b}) = \begin{bmatrix}
    b_+ & 0 \\ 0 & b_-
    \end{bmatrix}\,, 
\end{align}
where $a_\pm = \frac{1}{2}(1\pm r_{z,\textsc{a}})$ and $r_{z,\textsc{a}}$ is the $z-$component of the Bloch vector for qubit $A$ (resp. for $B$). Thermal states would correspond to $0<r_{z,j}<1$. In the uncoupled basis $\{\ket{g_{\textsc{a}}g_{\textsc{b}}},\ket{g_{\textsc{a}}e_{\textsc{b}}},\ket{e_{\textsc{a}}g_{\textsc{b}}},\ket{e_{\textsc{a}}e_{\textsc{b}}}\}$, a slightly tedious but direct computation shows that the relevant nonzero components of the $4\times 4$ matrix representation of $\hat{\rho}_\textsc{ab}$ are given by
\begin{equation}
    \begin{aligned}
    4{\rho}_{11} &= r_{z,\textsc{b}}\nu_{\textsc{b}} (r_{z,\textsc{a}} \cosh(2H_{\textsc{ab}}) \nu_{\textsc{a}}+\cos(2E_{\textsc{ab}}))+r_{z,\textsc{a}} \nu_{\textsc{a}}+1
    \,,\\
    4{\rho}_{14} &= -ir_{z,\textsc{b}}\nu_{\textsc{b}} e^{i \Omega  (\tau_{\textsc{a}}+\tau_{\textsc{b}})} (\sin(2E_{\textsc{ab}})-r_{z,\textsc{a}} i\sinh(2H_{\textsc{ab}}) \nu_{\textsc{a}}) \,,\\
    4\hat{\rho}_{22} &= -r_{z,\textsc{b}} \nu_{\textsc{b}} (r_{z,\textsc{a}} \cosh(2H_{\textsc{ab}}) \nu_{\textsc{a}}+\cos(2E_{\textsc{ab}}))+r_{z,\textsc{a}} \nu_{\textsc{a}}+1\,,\\
    4{\rho}_{23} &=  ir_{z,\textsc{b}}\nu_{\textsc{b}} e^{i \Omega  (\tau_{\textsc{a}}-\tau_{\textsc{b}})} (\sin(2E_{\textsc{ab}})-r_{z,\textsc{a}} i\sinh(2H_{\textsc{ab}}) \nu_{\textsc{a}})\,,\\
    %{\rho}_{32} &= -ir_{z,\textsc{b}}\nu_{\textsc{b}} e^{-i \Omega  (t_{\textsc{a}}-t_{\textsc{b}})} (r_{z,\textsc{a}} i\sinh(2H_{\textsc{ab}}) \nu_{\textsc{a}}+\sin(2E_{\textsc{ab}}))\,,\\
    4{\rho}_{33} &= -r_{z,\textsc{a}}  \nu_{\textsc{a}} (r_{z,\textsc{b}} \cosh(2H_{\textsc{ab}})\nu_{\textsc{b}}+1)+r_{z,\textsc{b}}\cos(2E_{\textsc{ab}}) \nu_{\textsc{b}}+1\,,
    %{\rho}_{41} &= ir_{z,\textsc{b}}\nu_{\textsc{b}} e^{-i \Omega  (t_{\textsc{a}}+t_{\textsc{b}})} (r_{z,\textsc{a}} i\sinh(2H_{\textsc{ab}}) \nu_{\textsc{a}}+\sin(2E_{\textsc{ab}}))\,,\\
    %{\rho}_{44} &= r_{z,\textsc{a}} \nu_{\textsc{a}} (r_{z,\textsc{b}} \cosh(2H_{\textsc{ab}}) \nu_{\textsc{b}}-1)-r_{z,\textsc{b}} \cos(2E_{\textsc{ab}}) \nu_{\textsc{b}}+1\,,
    \end{aligned}
    \label{eq: components}
\end{equation}
where $\nu_j = \exp(-2\mathsf{W}_{jj})$, with $W_{jk}$ being the smeared Wightman two-point functions,
\begin{align}
    \mathsf{W}_{jk}\coloneqq \mathsf{W}(f_j,f_k) = \braket{\hat\phi(f_j)\hat{\phi}(f_k)}\,.
\end{align}
%Other components (such as $\rho_{41}$) can be obtained by complex conjugation.

We can always decompose the smeared Wightman function $\mathsf{W}_{jk}$ into the smeared anti-commutator (\textit{Hadamard function}) $H_{jk}\coloneqq H(f_j,f_k) = \braket{\{ \hat\phi(f_j),\hat{\phi}(f_k)\}}$ and smeared commutator (\textit{causal propagator}) $E_{jk} \coloneqq E(f_j,f_k) = \braket{[\hat\phi(f_j),\hat{\phi}(f_k)]}$, i.e.,  $ \mathsf{W}_{jk} = \frac{1}{2} H_{jk} + \frac{i }{2} E_{jk}$. It is the non-vanishing of $H_{jk}$ that allows for the metric reconstruction proposal in \cite{Kempf2021replace,perche2021geometry}: for us, the form we want is the smeared version, which is given in \cite{tjoa2022modest}:
\begin{align}
    g_{\mu\nu}(\sx) &\approx {-\frac{1}{8\pi^2\delta^2}}\Bigr[\mathsf{W}(f_\epsilon,g_\epsilon)^{-1} - \mathsf{W}(f_\epsilon,g)^{-1} -\mathsf{W}(f,g_\epsilon)^{-1}+\mathsf{W}(f,g)^{-1}\Bigr]\,,
    \label{eq: metric-reconstruction}
\end{align}
where $f$ is centred around $\sx$, $\delta$ is the effective size of $f,g$, and $\epsilon$ is effective separation between the qubits. Both $\delta,\epsilon>0$ limit our ability to reconstruct the metric perfectly.

Eq.~\eqref{eq: components} tells us that we can, in principle, reconstruct the metric by quantum state tomography of $\hat{\rho}_{\textsc{ab}}$. One way to do this is to perform the joint measurement of Pauli operators (possibly by a third party, e.g., Alice and Bob's supervisor) post-interaction. For example:
\begin{equation}
\begin{aligned}
    \braket{\hat{\sigma}^x\otimes \hat{\sigma}^x} &= %\tr(\hat{\rho}_{\textsc{ab}}\hat{\sigma}^x\otimes\hat{\sigma}^x) 
    r_{z,\textsc{b}} \nu_\textsc{b} \sin ( \Omega \tau_{\textsc{b},0} ) \Bigr(\sin(2E_\textsc{ab}) \cos (\Omega \tau_{\textsc{a},0}) +r_{z,\textsc{a}}\nu_\textsc{a}  \sinh(2H_\textsc{ab}) \sin ( \Omega \tau_{\textsc{a},0} )\Bigr)\,,\\
    \braket{\hat{\sigma}^y\otimes \hat{\sigma}^y} &= %\tr(\hat{\rho}_{\textsc{ab}}\hat{\sigma}^y\otimes\hat{\sigma}^y) 
    r_{z,\textsc{b}}\nu_\textsc{b}\cos(\Omega\tau_{\textsc{b},0}) \Bigr(\sin(2E_\textsc{ab})\sin(\Omega\tau_{\textsc{a},0}) + r_{z,\textsc{a}}\nu_\textsc{a}\sinh(2H_\textsc{ab})\cos(\Omega\tau_{\textsc{a},0}) \Bigr)\,,\\
    \braket{\hat{\sigma}^z\otimes \hat{\sigma}^z} &= %\tr(\hat{\rho}_{\textsc{ab}}\hat{\sigma}^z\otimes\hat{\sigma}^z) 
    r_{z,\textsc{a}}r_{z,\textsc{b}}\nu_{\textsc{a}}\nu_\textsc{b}\cosh(2H_\textsc{ab})\,.
\end{aligned}
\end{equation}
It is therefore sufficient to recover the Wightman function from two out of the three non-local measurements above\footnote{Local measurements of the form $\braket{\hat{\sigma}^j\otimes\openone}$ or $\braket{\openone\otimes \hat{\sigma}^j}$ can at best reconstruct the causal propagator $E_{\textsc{ab}}$.}. 

Since we do not use perturbative series expansion, the calculation works for all values of coupling strengths of Alice and Bob's detectors. There are several interesting regimes:
\begin{itemize}[leftmargin=*]
    %\item If the initial state is maximally mixed ($r_{z,j}=0$) then the measurement statistics say nothing about the Wightman function.

    \item If either one of them are strongly coupled to the field ($\lambda_j\to\infty$), we have $\nu_{j}\to 0$ since $\mathsf{W}_{jj} \propto \lambda_j^2$: we lose all information about the Wightman function.

    \item If Alice is strongly coupled but Bob is weakly coupled, we have $\nu_\textsc{a}\approx 0,\nu_\textsc{b}\approx 1$. Then while $\braket{\hat{\sigma}^z\otimes \hat{\sigma}^z}$ gives nothing, we can get at least reconstruct the causal structure from $E_{\textsc{ab}}$ via $\braket{\hat{\sigma}^x\otimes \hat{\sigma}^x}$. However, since $E_{\textsc{ab}}\propto \lambda_{\textsc{a}}\lambda_{\textsc{b}}$, $\sin(E_{\textsc{ab}})$ becomes highly oscillatory in the regime $\lambda_{\textsc{a}}\lambda_\textsc{b}\gg 1$ (i.e., a small change in $\lambda_{\textsc{a}}\lambda_{\textsc{b}}$ leads to a large change in $\sin (E_{\textsc{ab}})$.

    \item If the coupling of both parties are too weak, it is also very difficult to reconstruct the metric because $H_\textsc{ab},E_{\textsc{ab}}\propto \lambda_\textsc{a}\lambda_\textsc{b}$, since the leading contribution to the Pauli measurements is $\mathcal{O}(1)$ and does not contain $H_{\textsc{ab}}$. Reconstruction of $E_{\textsc{ab}}$ is still possible so long as Alice and Bob are not spacelike-separated.
    
    %However, this is not so straightforward as $E_{\textsc{ab}}\sim \lambda_{\textsc{a}}\lambda_{\textsc{b}}$ (linear in coupling strengths of each), so one has to find a balance in order to make sure that $\sin(2E_{\textsc{ab}})$ is not close to zero. However, by construction we cannot make Alice couple to the field in a much weaker strength than Bob \eri{Why? --- because we \textit{assumed} that $\hat{U} = \hat{U}_\textsc{b}\hat{U}_\textsc{a}$.}: when Alice's coupling is too weak compared to Bob, what ends up happening is that we lose the ability to measure the Hadamard part, and we can only recover the commutator $E_{\textsc{ab}}$. 

    %\item In general, it is bad strategy to implement weak coupling for both because $E_{\textsc{ab}},H_{\textsc{ab}}\to 0$ in the sine/cosine. One option is to fix the coupling strength and then by taking ``inverse cosine/sine'' we divide by the coupling strengths. 

    %\item So long as we consider only Hadamard states with the right UV-singularity behaviour, we always need $H_{\textsc{ab}}$ for specifying the metric. In general, we do not need to measure along the null directions --- but if we \textit{only} measure along the null directions, we effectively only measure the field commutator. This works for conformally flat spacetimes. 
\end{itemize}
These show that the coupling strengths (keeping everything fixed) must be sufficiently controlled to extract $\mathsf{W}_{\textsc{ab}}$. Thus we fully realize the notion that \textit{strong coupling leads to poor statistics}, and strong coupling is essentially high-energy/short-distance regime.

If the fundamental limitations of our measurements are quantum-mechanical in nature, of which the model we presented is one concrete realization, then the spacetime is already ``fuzzy'' way above the Planck scale. %\footnote{It is important to note that in practice,  the \textit{most precise} way to reconstruct the metric may very well be classical methods. This is analogous to how \textit{quantum clocks} may not be the \textit{best clock} as far as precision is concerned, but quantum clocks test fundamental limits of time-keeping \cite{meier2023fundamental}. Here we are also concerned with fundamental limits of metric reconstruction below some UV scale.}. 
This is a simple consequence of the fact that \textit{the size of the interaction region matters}: one cannot take the spacetime smearing function $f_j\sim \delta^4(\sx-\sx_j)$, i.e., localized in \textit{both} space \textit{and} time (equivalent to setting $\delta\to 0$ in \eqref{eq: metric-reconstruction}) due to the distributional nature of unsmeared two-point functions. Since we are already localizing the temporal profile sharply in \eqref{eq: delta-interaction}, the spatial profile must be finite in extent to avoid UV divergences in $\mathsf{W}_{jk}$. Therefore, even though mathematically there is no problem in taking formal limit of the unsmeared Wightman function to give $g_{ab}(\sx)\sim \lim_{\sx'\to\sx}\mathsf{W}(\sx,\sx')^{-1}$ \cite{Kempf2021replace,perche2021geometry,tjoa2022modest}, in practice any quantum-optical reconstruction of the metric from the correlators will have \textit{finite resolutions} given by the effective size of the spatial profile --- the size of the detector is nothing but the physical ultraviolet (UV) cutoff. {This ``fuzzy spacetime'' philosophy using UDW detector  construction was first concretized in \cite{perche2021geometry}, primarily motivated by the idea that beyond certain regimes classical rulers and clocks may not be available, and quantum-mechanical detectors and correlators should replace them.}

%Finally, it is interesting to see how the UDW model 

The result is also affected by some thermodynamic considerations. For example, in the high-temperature regime of Alice and Bob's state we have that $r_{z,j}\ll 1$ (which corresponds to $\beta_j\Omega\ll 1$). In this limit we see that the Pauli measurements \textit{lose all information} about the two-point functions and hence the metric reconstruction does not work. Interestingly, if Bob's initial state is very close to the ground state (very low-temperature), then the Pauli measurements can at least reconstruct the light cone from $E_{\textsc{ab}}$ (hence the metric up to a conformal factor \cite{Kempf2021replace,hawking2023large}) even if Alice's qubit is very hot so long as Alice and Bob are not spacelike-separated. The asymmetry between Alice and Bob is due to the fact that Alice couples to the field first before Bob (with respect to some global time). Note that we have already assumed idealized measurements; we also ignored the work cost from turning on the interactions and from the work associated with the renormalized stress-energy tensor \cite{Landulfo2021cost}. We expect that including any extra factors due to imperfections in measurements, among others things, will only make things worse. In this sense, our results is loosely analogous to Dyson's argument about detectability of gravitons \cite{dyson2013graviton}: that is, any ``literal'' implementations will not work easily and some workaround is needed.

{The most interesting result for us is that this fuzziness in metric reconstruction} has a dual holographic interpretation in terms of \textit{modest holography} \cite{tjoa2022modest}, which in turn is based on the bulk-to-boundary correspondence between QFT living in the bulk spacetime $\M$ and its null boundary (the future null infinity $\skri^+$): 
\begin{align}
    g_{\mu\nu}(\sx) &\approx {\frac{-1}{8\pi^2\delta^2}}\Bigr[\mathsf{W}_\skri(\psi_{f_\epsilon},\psi_{g_\epsilon})^{-1} - \mathsf{W}_\skri(\psi_{f_\epsilon},\psi_{g})^{-1}-\mathsf{W}_\skri(\psi_{f},\psi_{g_\epsilon})^{-1}+\mathsf{W}_\skri(\psi_{f},\psi_{g})^{-1}\Bigr]\,.
    \label{eq: metric-reconstruction-boundary}
\end{align}
The difference with \eqref{eq: metric-reconstruction} is that $\mathsf{W}_\skri(\psi_{f},\psi_{g})$ are two-point functions for the boundary scalar field theory living at $\skri^+$. The boundary smearing functions $\psi_f$ are obtained by ``projection to the boundary'', i.e., $\psi_f \equiv \lim_{\skri^+}(r Ef)(u,r,x^A)$, where $(u,r,x^A)$ are the Bondi-Sachs coordinates adapted for asymptotically flat spacetimes near $\skri^+$. 

The dual picture is based on the fact that $\mathsf{W}_\skri(\psi_{f},\psi_{g})$ has universal structure \cite{Dappiaggi2005rigorous-holo,Dappiaggi2009Unruhstate,Dappiaggi2008cosmological,dappiaggi2015hadamard}
\begin{align}
    \mathsf{W}_\skri(\psi_f,\psi_g)
    &= -\frac{1}{\pi}\lim_{\varepsilon\to 0}\int {\dd \gamma_{S^2}}\dd u\dd u'\frac{\psi_f(u,x^A)\psi_g(u',x^A)}{(u-u'-i \varepsilon)^2}\,.
    \label{eq: boundary-Wightman}
\end{align}
This in fact coincides with the smeared Wightman two-point function associated to \textit{pointlike} qubits with \textit{finite temporal switching function}, in contrast to delta-switching in \eqref{eq: delta-interaction}. In other words, we have the following dual interpretation: the metric reconstruction using short-time, finite-sized qubits with spacetime profile $f,g$ (\textit{c.f.} \eqref{eq: delta-interaction}) in the bulk $\M$ is holographically equivalent to metric reconstruction using \textit{finite-time, pointlike\footnote{The pointlikeness near the boundary arises due to essentially taking large-$r$ expansion in Bondi-Sachs coordinates: any spatially finite objects appear pointlike from infinity.}} qubits (possibly with $\Omega=0$)  with \textit{temporal profile} $\psi_f,\psi_g$ from the boundary $\skri^+$. It is noteworthy that the boundary scalar field is technically a \textit{generalized free field}, since it has no equation of motion.  %The affine parameter $u$ takes the role of the proper times.

There is one additional interesting observation we can make from the RHS of \eqref{eq: boundary-Wightman}: it can be exactly identified with the pullback of the Wightman two-point function in \textit{Minkowski spacetime} associated to pointlike qubits, given by (see, e.g., \cite{tjoa2021harvesting,pozas2015harvesting,smith2016topology})
\begin{align}
    \mathsf{W}_{\text{flat}}(\chi_f,\chi_g) &\sim -\frac{1}{\pi}\lim_{\varepsilon\to 0}\int \frac{\dd\tau\,\dd\tau'\,\chi_f(\tau)\chi_g(\tau')}{(\tau-\tau'-i\varepsilon)^2} %\notag\\&
    \sim -\frac{1}{\pi}\lim_{\varepsilon\to 0}\int \frac{\dd\tau\,\dd\tau'\,\chi_f(\tau)\chi_f(\tau')}{(\tau-\tau'-i\varepsilon)^2-L^2}\,,
    \label{eq: flat}
\end{align}
with temporal switching functions $\chi_f(\tau) \coloneqq \psi_f$. In passing to the second line we have used the fact that $L$ is an effective (small) separation between center-of-mass of $f$ and $g$  \cite{tjoa2022modest}. This means that all the information about the metric can be equivalently encoded in non-trivial modification to the temporal profile in flat spacetime. Notice that from the perspective of Eq.~\eqref{eq: flat} (and also Eq.~\eqref{eq: boundary-Wightman}), the fuzziness of spacetime appears for a completely different reason: it has to do with the UV-divergent nature of \textit{temporal localization} for pointlike detectors. Therefore, the holographic duality says that (1) information in the bulk spatial profile is encoded in the boundary temporal profile according to Eq.~\eqref{eq: boundary-Wightman}; (2) at the detector level, the duality admits a \textit{flat spacetime} interpretation according to  Eq.~\eqref{eq: flat}. {Note that the modest holography proposal \textit{necessitates} the fuzziness \textit{even in principle}.}% The fuller details of the holographic reconstruction using particle detector models will be given elsewhere \cite{tjoa2023holography-detector}.

In summary, we construct a non-perturbative detector-field model that imposes natural quantum-optical limitations to metric reconstruction that appear way above the Planck scale. We argue that these limitations realize the principle that high energy/strong coupling leads to poor statistics. In particular, we show that the model also admits dual holographic interpretation using bulk-to-boundary correspondence, which forbids strict localization in spacetime if one were to use quantum field correlators to reconstruct the spacetime geometry. The fuzziness of spacetime in this context is closely tied to the quality of statistics of the measurements imposed on the quantum-mechanical detectors used to probe the geometry.

\noindent \textbf{Acknowledgments.} E.T. thanks Eduardo Mart\'in-Mart\'inez and Tales Rick Perche for useful discussions. This work is supported in part by the Natural Sciences and Engineering Council of Canada (NSERC). Research at the Institute for Quantum Computing is supported in part by the Government of Canada through the Department of Innovation, Science and Economic Development Canada and by the Province of Ontario through the Ministry of Colleges and Universities. 
Institute for Quantum Computing and the University of Waterloo are situated on the Haldimand Tract, land that was promised to the Haudenosaunee of the Six Nations of the Grand River, and is within the territory of the Neutral, Anishnawbe, and Haudenosaunee peoples.

\bibliography{ref}
\bibliographystyle{unsrt}

\end{document}